\documentclass[aps,prl,twocolumn,superscriptaddress, 10pt]{revtex4-1}
\usepackage{graphics,graphicx, array, afterpage}
\usepackage{qcircuit,amsmath}
\usepackage{grffile}
\usepackage[export]{adjustbox}
\usepackage{lineno}
\usepackage{xcolor}
\usepackage{longtable}
\usepackage{braket}
\usepackage{multirow}
\usepackage{enumerate}
\usepackage{hyperref}

\usepackage{amsmath}
\usepackage{amssymb}
\usepackage{amsthm}

\newcommand{\Gate}[1]{\textsc{#1}}
\newcommand{\cnotgate}{\Gate{cnot}} 
\newcommand{\swapgate}{\Gate{swap}} 
\newcommand{\hgate}{\Gate{h}}
\newcommand{\idgate}{\Gate{i}}
\newcommand{\xgate}{\Gate{x}}
\newcommand{\ygate}{\Gate{y}}
\newcommand{\zgate}{\Gate{z}}
\newcommand{\pgate}{\Gate{s}} 

\newcommand{\mgate}{\Gate{m}} 
 
\newcommand{\majfivefid}{0.9429}
\newcommand{\majfivestd}{0.0011}
\newcommand{\majeightfid}{0.9280}
\newcommand{\majeightstd}{0.0013}
\newcommand{\mbfid}{0.8743}
\newcommand{\mbstd}{0.0035}
\newcommand{\hxzhfid}{0.8460}
\newcommand{\hxzhstd}{0.0053}
\newcommand{\slsbsixfid}{0.7984}
\newcommand{\slsbsixstd}{0.0047}

\newcommand{\maj}[1]{\mathrm{MAJ}{#1}}
\newcommand{\slsb}[1]{\mathrm{SLSB}{#1}}
\newcommand{\ip}[1]{\mathrm{IP}{#1}}
\newcommand{\flip}{\mathsf{flip}}
\newcommand{\reset}{\mathsf{reset}}

\newcommand{\fig}[1]{\hyperref[fig:#1]{Fig.~\ref*{fig:#1}}}
\newcommand{\tab}[1]{\hyperref[tab:#1]{Table~\ref*{tab:#1}}}
\newcommand{\lem}[1]{\hyperref[lem:#1]{Lemma~\ref*{lem:#1}}}
\newcommand{\thm}[1]{\hyperref[thm:#1]{Theorem~\ref*{thm:#1}}}

\begin{document}

\title{Quantum advantage for computations with limited space}

\author{Dmitri Maslov}
\affiliation{IBM Quantum, IBM T.J. Watson Research Center, Yorktown Heights, NY 10598, USA}
\author{Jin-Sung Kim}
\affiliation{IBM Quantum, Almaden Research Center, San Jose, CA 95120, USA}
\author{Sergey Bravyi}
\affiliation{IBM Quantum, IBM T.J. Watson Research Center, Yorktown Heights, NY 10598, USA}
\author{Theodore J. Yoder}
\affiliation{IBM Quantum, IBM T.J. Watson Research Center, Yorktown Heights, NY 10598, USA}
\author{Sarah Sheldon}
\affiliation{IBM Quantum, Almaden Research Center, San Jose, CA 95120, USA}

\date{\today}


\begin{abstract}
Quantum computations promise the ability to solve problems intractable in the classical setting \cite{nielsen2011quantum}.  Restricting the types of computations considered often allows to establish a provable theoretical advantage by quantum computations \cite{bell1964einstein, deutsch1992rapid, bernstein1997quantum, grover1996fast, trotter1959product, shor1999polynomial, bravyi2018quantum}, and later demonstrate it experimentally \cite{aspect1982experimental, debnath2016demonstration, figgatt2017complete, vandersypen2001experimental}.  In this paper, we consider space-restricted computations, where input is a read-only memory and only one (qu)bit can be computed on.  We show that $n$-bit symmetric Boolean functions can be implemented exactly through the use of quantum signal processing \cite{low2017optimal} as restricted space quantum computations using $O(n^2)$ gates, but some of them may only be evaluated with probability $1/2 \,{+}\, O(n/\sqrt{2}^n)$ by analogously defined classical computations.  We experimentally demonstrate computations of $3$-, $4$-, $5$-, and $6$-bit symmetric Boolean functions by quantum circuits, leveraging custom two-qubit gates, with algorithmic success probability exceeding the best possible classically.  This establishes and experimentally verifies a different kind of quantum advantage---one where
quantum scrap space is more valuable than analogous classical space---and calls for an in-depth
exploration of space-time tradeoffs in quantum circuits.
\end{abstract}

\maketitle 

\section{Introduction}
Quantum computations are studied for their potential to offer an advantage over regular classical computations.  The extent and provability of such advantage depend on the computational model selected.  A simple example of a computational model can be a game.  Consider the CHSH game \cite{clauser1969proposed} (Bell's inequality \cite{bell1964einstein}), where the two players Alice and Bob are given random Boolean inputs $s$ and $t$ and are required to come up with the bits $a$ and $b$, respectively, and using no communication, such that $s{\wedge}t \,{=}\, a {\oplus}b$.  The best classical probability of winning this game, $\frac{3}{4}$, can be improved to $\frac{2+\sqrt{2}}{4}$ with the use of a quantum computer.  While this gap allows to experimentally demonstrate quantumness, there is very little quantum computation involved, and Bell's inequality can be attributed to the property of quantum states rather than computations.  A second model studies computations with black boxes.  It allows proving computational complexity separations for a set of problems such as distinguishing constant and balanced functions (Deutsch-Jozsa \cite{deutsch1992rapid}), discovering a hidden linear reversible function (Bernstein–Vazirani \cite{bernstein1997quantum}), and finding a satisfying assignment (Grover \cite{grover1996fast}).  While no practical utility is known for the first two problems/algorithms, Grover's search can be employed to find a satisfying assignment for some difficult to invert but efficiently computable function.   However, given a mere quadratic quantum speedup in a model that does not account for the cost of implementing the oracle, practical utilization of Grover's search is likely far in the future.  A third computational model studies white box computations, and allows superpolynomial advantage for solving problems such as Hamiltonian dynamics simulation \cite{trotter1959product, suzuki1976generalized, low2017optimal} and discrete logarithm over Abelian groups (including Shor's integer factoring \cite{shor1999polynomial}).  In this case, separations are not established formally, although believed to hold, and a quantum computer capable of outperforming a classical computer will likely need to be large---about $70$ qubits and 650,000 gates in some of the shortest known quantum circuits solving a computational problem that is believed to be intractable for classical hardware \cite{nam2019low} (the resource counts assume perfect physical-level quantum computer, and are higher in the fault-tolerant scenario).  Finally, a provable quantum advantage was established for the parallel model of computation.  It was shown that parallel quantum algorithms can solve certain computational problems in constant time, whereas the best possible classical algorithm takes time growing at least logarithmically with the input size~\cite{bravyi2018quantum,bravyi2020quantum,le2019average,coudron2018trading,watts2019exponential,grier2020interactive}.  It remains to be seen whether this type of advantage can be demonstrated experimentally with near-term devices due to the large number of qubits required.

Here we study a simple computational model that allows to both establish a provable separation between classical and quantum computational models and validate it experimentally.  Our model is designed to highlight the superiority of quantum computational space, resulting in a different type of advantage compared to those examples highlighted in the previous paragraph.  A related space advantage should be possible to exploit to improve computations beyond those explicitly discussed in this paper. 

\section{Theory}
Formally, we consider classical and quantum circuits where input (also called primary input to distinguish from the constant qubit called the computational space) is a read-only memory (input cannot be written on), and the computational space is restricted to $s$ bits.  In the classical case, computations proceed by arbitrary $s{+}1$-input $s$-output Boolean functions/gates $g$, where exactly one bit of the input to $g$ is from the primary input, and all outputs are computational bits.  For $s{=}1$ this means $2$-input $1$-output Boolean gates, being the staple gate library for classical computations.  The closest analog to such transformations in the quantum world is the controlled-$U$ gates, where the unitary operation $U$ is applied to the computational register and controlled by a primary input.  We call this model limited-space computation.

The set of functions uncomputable by $1$-bit limited-space classical computations includes symmetric functions with nontrivial Fourier spectra (equivalently, those that cannot be written as fixed polarity Reed-Muller expression with degree $0$, $1$, and $n$ terms only).  This implies that most symmetric functions may not be computed classically in this model.  However, they can be computed by a quantum circuit with $O(n^2)$ entangling gates and $1$ qubit of computational space, as discussed later.  Other than symmetric Boolean functions, polynomial-size $1$-qubit limited-space quantum computations include at least those functions in the ${\text NC}^1$ class, such as Boolean components of the integer addition, integer multiplication, and matrix determinant \cite{ablayev2005computational}, as well as all linear combinations $f(x)\,{\oplus}\,g(y)$ where $f$ and $g$ are polynomial-size computable; most of these functions are uncomputable by $1$-bit limited-space classical computations.  

When the computational space is increased to $2$ bits, the classical model can compute any Boolean function (e.g., by Disjunctive Normal Form), although the circuit complexity may be high. For example, assuming ${\text NC}^1\,{\neq}\, {\text ACC}$, Majority cannot be implemented exactly or with probability greater than $\frac{7}{8}$ using a polynomial sized circuit and just $2$ computational bits \cite{ablayev2005computational}. When the computational space is increased to $3$ bits, Barrington’s theorem \cite{barrington1989bounded} promises a polynomial-sized circuit, but large exponents seem inevitable.  For instance, the best circuit for Majority with $3$ computational bits still consists of $O(n^{5.42})$ gates (see Methods).  In this paper, we show that quantum computers with a single computational qubit can compute all symmetric Boolean functions exactly with circuits of size just $O(n^2)$, demonstrating advantage against classical circuits even if they are allowed up to $3$ bits of computational space.

With perfect quantum computers, we would be able to demonstrate that the quantum computer always succeeds at computing those functions uncomputable by the classical $1$-bit limited-space circuits.  Unfortunately, current quantum computers are noisy and sometimes fail.  This failure is often modeled probabilistically.  To demonstrate quantum advantage using noisy quantum computers over (perfect) classical computers in an experiment, it would be fair to arm classical computations with free access to randomness.  Specifically, we allow the classical computer to randomly select a limited-space circuit to run or, equivalently, replace Boolean gates $g(x_i,s)$ in it with Boolean gates $g(x_i,s,r)$, where $r$ is a random number.  We furthermore allow the classical limited-space computer to evaluate functions with probability $p$, which is equal to the normalized Hamming distance between truth vectors of the computable and desired functions.  The value $p$ for classical computations is thus analogous to ASP (Algorithmic Success Probability) in quantum computations.  Computational machinery that achieves ASP above the maximal classical value $p$ performs a computation unreachable by classical means and is thereby super-classical.  Here we demonstrate a selection of experiments that achieve this.

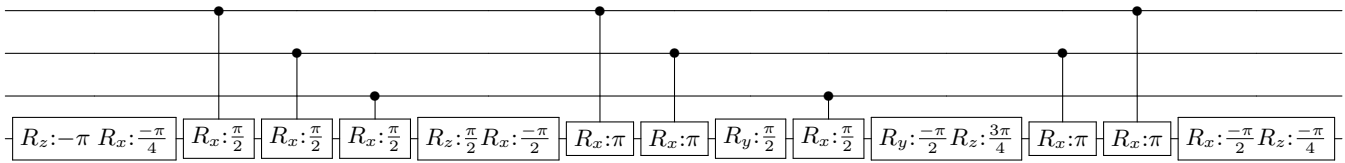
\begin{figure*}[t]
\Qcircuit @C=0.3em @R=0em @!R {
& \qw 	& \ctrl{3} & \qw 	& \qw 		& \qw 		& \ctrl{3}  & \qw  & \qw 	& \qw & \qw & \qw & \ctrl{3} & \qw & \qw \\
& \qw 		& \qw 		& \ctrl{2} 	& \qw 		& \qw 		& \qw   & \ctrl{2}  & \qw & \qw & \qw & \ctrl{2}	& \qw & \qw & \qw \\
& \qw 		& \qw 		& \qw 	& \ctrl{1}		& \qw 		& \qw   & \qw 		& \qw 		& \ctrl{1} & \qw 		& \qw 		& \qw & \qw & \qw \\
& \gate{R_z{:}{-}\pi \; R_x{:}\frac{-\pi}{4}}  & \gate{R_x{:}\frac{\pi}{2}}	& \gate{R_x{:}\frac{\pi}{2}}	& \gate{R_x{:}\frac{\pi}{2}}	& \gate{R_z{:}\frac{\pi}{2} R_x{:}\frac{-\pi}{2}}	& \gate{R_x{:}{\pi}}	& \gate{R_x{:}{\pi}}	& \gate{R_y{:}\frac{\pi}{2}} & \gate{R_x{:}\frac{\pi}{2}} & \gate{R_y{:}\frac{-\pi}{2}R_z{:}\frac{3\pi}{4}} & \gate{R_x{:}{\pi}} & \gate{R_x{:}{\pi}} & \gate{R_x{:}\frac{-\pi}{2}R_z{:}\frac{-\pi}{4}} & \qw
}
\caption{\label{fig:circ} True $\slsb{3}$ with $8$ entangling gates, obtained using signal processing technique and local optimization. The gates used are axial rotations $R_{\{x,y,z\}}{:}\theta$ by the angle $\theta$ and their controlled versions.}
\end{figure*}

\begin{figure}[t]
\hspace{5mm}\Qcircuit @C=0.4em @R=0.1em @!R {
\lstick{x_1} & \qw & \qw & \qw 	& \ctrl{5} 	& \qw 	& \qw   & \qw & \qw \\
\lstick{x_2} & \qw & \qw & \ctrl{4} 	& \qw   & \ctrl{4} & \qw  & \qw & \qw \\
\lstick{x_3} & \qw & \ctrl{3} & \qw 	& \qw 	& \qw	& \ctrl{3}  & \qw & \qw  \\
\lstick{\ldots} &&&&&&& \\	
\lstick{x_n} & \ctrl{1} & \qw	& \qw 	& \qw	& \qw	& \qw   & \ctrl{1} 	& \qw \\
\lstick{\ket{b}} & \gate{\hgate\xgate} & \gate{\hgate\xgate}	& \gate{\hgate\xgate}	& \gate{\zgate}	& \gate{\hgate}	& \gate{\hgate}	& \gate{\hgate}	& \qw
}
\caption{\label{fig:circ2} Relative-phase $\slsb{n}$ using $2n{-}1$ entangling gates.  The gates used are the controlled versions of the Clifford gate $\hgate\xgate$, Pauli-$Z$ gate, and the Hadamard gate $\hgate$.}
\end{figure}
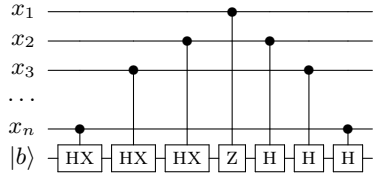

The simplest function not computable in the $1$-bit limited-space classical model is $\slsb{3}(x_1,x_2,x_3)=x_1x_2 {\oplus} x_2x_3 {\oplus} x_1x_3$.  In general, $\slsb{n}$ is defined as the value of the Second Least Significant Bit of the input weight $|x|$.  The maximal classical probability $p$ of computing $\slsb{3}$ using limited-space computations is $\frac{7}{8}\,{=}\,0.875$, meaning the truth vector distance to a computable function is $1$.  We developed two quantum circuits to compute $\slsb{3}$, one with $5$ entangling gates (\fig{circ2}) and one with $8$ entangling gates (\fig{circ}), achieved with the use of Quantum Signal Processing (QSP).  The quantum computer ASPs are \majfivefid{}$\pm$\majfivestd{} and \majeightfid{}$\pm$\majeightstd{}, respectively.  For $4$ bits, the function $\slsb{4}$ achieves the minimal among maximal classical values $p\,{=}\,\frac{13}{16}\,{=}\,0.8125$ across all symmetric Boolean functions.  We developed a quantum circuit with $7$ entangling gates, \fig{circ2}, that maps into a quantum circuit with $13$ entangling gates over the experiment, due to the requirement to use two $\swapgate$ gates.  The measured ASP is \mbfid{}$\pm$\mbstd{}.  For $5$ bits, the function $\slsb{5}$ is most difficult to approximate classically, with the threshold of $\frac{23}{32}\,{=}\,0.71875$; we achieved quantum ASP of \hxzhfid{}$\pm$\hxzhstd{} by a quantum circuit with 9 entangling gates (21 in the experiment), \fig{circ2}.  For $6$ bits, the most difficult function is $\slsb{6}$, featuring the threshold value $\frac{43}{64}\, {=}\,0.671875$; we implemented it with fidelity $\slsbsixfid{}{\pm}\slsbsixstd{}$ over quantum circuit with $11$ gates ($29$ in the experiment), \fig{circ2}.  In each of these experiments, we beat the classical threshold, thus demonstrating a quantum advantage.  

For arbitrary $n$, $\slsb{n}$ as well as any symmetric Boolean function, can be computed using $O(n^2)$ entangling gates by a quantum limited-space circuit, constructed using QSP.  $\slsb{n}$ may furthermore be computed by a specialized circuit using $2n{-}1$ gates (see \fig{circ2}), showing that QSP gives a loose upper bound.  The classical probability $p$ of evaluating $\slsb{n}$ correctly within the limited-space computational model approaches the theoretical minimum of $\frac{1}{2}$ exponentially fast, namely, $p\,{\le}\,1/2 {+} O(n/\sqrt{2}^n)$.  This presents an opportunity to demonstrate larger quantum advantage with a higher number of qubits.  Formal proofs of the above statements are deferred to the Methods section.

Our goal is the construction of a quantum circuit implementation of the $n$-bit Boolean function $f(x)$, expressed by an $n{+}1$-qubit unitary $U:\ket{x}\ket{b}\rightarrow e^{i\theta(x,b)}\ket{x}\ket{b{\oplus}f(x)}$ for some real-valued function $\theta(x,b)$.  In the 1-qubit limited-space model we may write $U\,{=}\,\sum_x\ket{x}\bra{x}\otimes U(x)$, where $U(x)$ is the product of single-qubit gates, each controlled by a single qubit of the input register $\ket{x}$. We show in the Methods section that the simplest implementation of $U$ in which $\theta(x,b)$ is constant and $U(x)\,{=}\,e^{i\theta(x,b)}X^{f(x)}$ is impossible. The closest we can get to such a phaseless implementation is $U(x)\,{=}\,(iX)^{f(x)}$, which we call a \emph{true} implementation.  Any other case we regard as a \emph{relative phase} implementation.  Note that both true and relative phase implementations faithfully compute $f(x)$ upon measurement in the computational basis.  An advantage of true implementation comes from the ability to remove the phase entirely through introducing a new ancilla qubit. 

Our structured approach to computing symmetric Boolean functions $f(x)$ makes use of QSP \cite{low2017optimal}.  Suppose that we only access the input bits with a unitary $S\,{=}\,\sum_{x}\ket{x}\bra{x}{\otimes} R_x\left(\phi_x\right)$, where $R_P(\chi)=\cos(\chi/2)I\,{-}\,i\sin(\chi/2)P$ is a single-qubit rotation for any Pauli operator $P\,{\in}\,\{X,Y,Z\}$.  Letting $\phi_x\,{=}\,\Delta|x|{-}\delta$ for real parameters $\delta$ and $\Delta$, it is clear that we can implement $S$ with $n$ controlled-$R_x(\Delta)$ gates and an $R_x(\delta)$ gate.  QSP is a method to create $U$ using the $S$ operation several (say, $L$) times, interspersed with single-qubit gates on the computational qubit.  In the simplest case, sufficient for our purposes, these additional single-qubit gates are $Z$-rotations $R_z(\xi)$.  To be more concrete, suppose we write $U(x)\,{=}\,A(x)I{+}iB(x)X{+}iC(x)Y{+}iD(x)Z$ for real-valued functions $A$, $B$, $C$, and $D$. Provided that these functions satisfy $A^2{+}B^2{+}C^2{+}D^2\,{=}\,1$ and have certain symmetries, QSP guarantees the existence of angles $\xi_j$ such that $U\,{=}\,R_z(\xi_0)\prod_{j=1}^LR_z(\xi_j)SR^\dag_Z(\xi_j)$ and gives an efficient method to find these angles \cite{low2016methodology,haah2019product}.  See the Methods section for more details.

To compute a symmetric function of $n$ bits, we choose $A(x)\,{=}\,1$ when $f(x)\,{=}\,0$ and $B(x)\,{=}\,1$ when $f(x)\,{=}\,1$.  These constraints are satisfiable for $L\,{=}\,4n{+}1$ uses of $S$ (see Methods).  Since each instance of $S$ uses $n{+}1$ gates, the total gate-complexity of this approach is $O(n^2)$. For certain functions $f$, symmetries and circuit simplifications can reduce this gate count.  For instance, the QSP approach calculates true $\slsb{3}\,{=}\,\maj{3}$ using $9$ entangling gates, and a simple gate merging simplification reduces their number to $8$, \fig{circ}(a). $\maj{n}$, the majority function, evaluates to one iff more than half the inputs equal one.  The true $5$-bit majority $\maj{5}$ implementation by QSP takes $25$ entangling gates.  For more general Boolean functions that lack the symmetry present in $\maj{n}$, gate counts are larger. For instance, an unoptimized QSP circuit for the true implementation of $\slsb{4}$ function, which operates over fewer bits than $\maj{5}$, takes $52$ entangling gates.  In contrast, a relative-phase implementation constructed directly has only $7$ entangling gates, \fig{circ2}. 


\begin{figure*}[t]
\includegraphics[width=180mm]{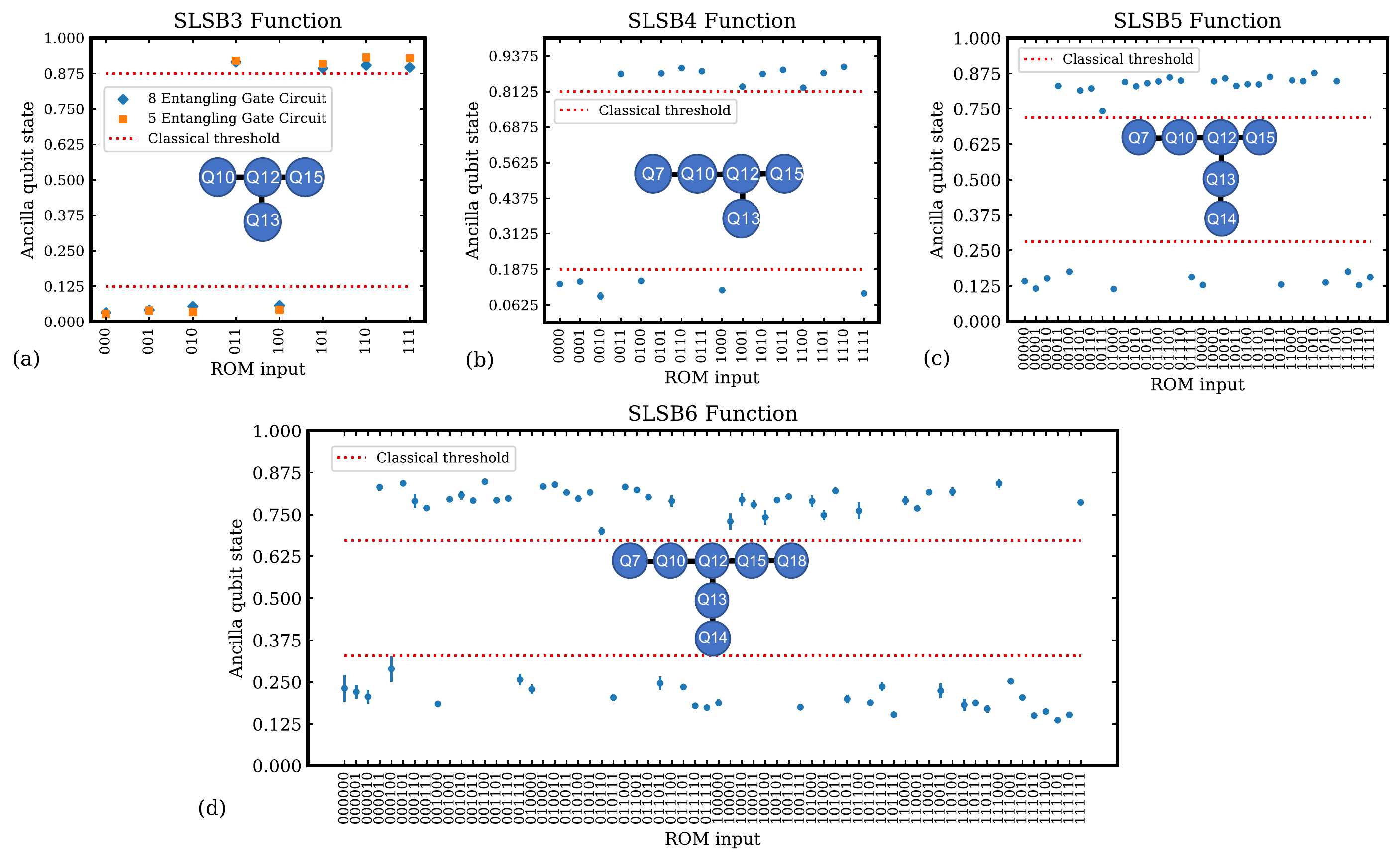}
\caption{Ancilla qubit (Q$12$) excited state population for each input ROM state combination for the (a) $\slsb{3}$, (b) $\slsb{4}$ (c) $\slsb{5}$ function and (d) $\slsb{6}$. The $5$ and $8$ entangling gate $\slsb{3}$ circuits achieve the ASP of \majfivefid{}$\pm$\majfivestd{} and \majeightfid{}$\pm$\majeightstd{}, while the $\slsb{4}$ circuit achieves the ASP of \mbfid{}$\pm$\mbstd{}, the $\slsb{5}$ circuit achieves the ASP of \hxzhfid{}$\pm$\hxzhstd{}, and the $\slsb{6}$ circuit achieves the ASP of \slsbsixfid{}$\pm$\slsbsixstd{}. Experimental error bars for each function are smaller than the plot marker in most cases.  The maximal classical ASP is illustrated with the red dotted lines.  In addition, the qubit layout for each function is displayed in the inset of each plot.}
\label{fig:maj3_mb4_data}
\end{figure*}

\section{Experiment}
We implement the circuits for $\slsb{3}$, $\slsb{4}$, $\slsb{5}$ and $\slsb{6}$ functions in Qiskit \cite{Qiskit}, an open-source quantum software development platform, and execute them on four to seven fixed-frequency superconducting transmon qubits on \textit{ibmq\_berlin}, a $27$-qubit heavy hexagonal lattice device \cite{chamberland2019} (\fig{berlin} in the Methods section). For each experiment, Q$12$ is utilized as the computational qubit and a subset of Q$7$, Q$10$, Q$13$, Q$14$, Q$15$, and Q$18$ is used as the read-only memory (ROM) inputs, depending on the circuit (\fig{maj3_mb4_data} insets). 


To execute these circuits efficiently, we calibrate a custom two-qubit entangling gate and add single-qubit rotations to implement the following gates: controlled-$R_{x}(\pi/2)$ and controlled-$\hgate\xgate$.  These gates are locally equivalent to a $ZX_{\pi/4}$ rotation, so while these gates can be implemented using two $\cnotgate$ gates, the circuit length and overall performance is improved by calibrating and implementing $ZX_{\pi/4}$ directly.  In addition, we implement the two-qubit gates that are locally equivalent to  $\cnotgate$, controlled-$\hgate$ and controlled-$\zgate$, using the single-qubit and $\cnotgate$ gates included in Qiskit's gate set. These standard gates are automatically calibrated daily. 

To calibrate the custom entangling gates, we utilize Qiskit Pulse \cite{alexander2020, garion2020}, a Qiskit module that allows the user to bypass the gate abstraction layer and implement controls directly at the microwave pulse level.  Entangling gates between coupled qubits are achieved using an echoed cross-resonance (CR) microwave pulse in which a Gaussian-square shaped positive CR tone is applied to the control qubit at the target qubit's resonant frequency \cite{rigetti2010,chow2011}.  A subsequent $\pi$ pulse and negative CR tone applied to the control qubit echo away unwanted Hamiltonian terms \cite{sheldon2016}.  The resulting interaction is primarily a $ZX$ term which provides a conditional $X$ rotation on the target qubit depending on the state of the control qubit. 

A single echoed $ZX_{\pi/4}$ CR tone must be calibrated per ROM input coupled to the computational qubit Q$12$ to implement the custom gates controlled-$R_{x}(\pi/2)$ and controlled-$\hgate\xgate$.  Using Qiskit Pulse, we modify the length of the standard $\cnotgate$ gate CR tone to roughly half the original duration, then recalibrate the amplitude to perform the $ZX_{\pi/4}$ rotation. In addition, a resonant $2\pi$ rotary echo is calibrated and applied to the target qubit simultaneously with the CR tone in order to minimize cross-talk between the target and spectator qubits \cite{sundaresan2020}.  After calibration, we assess the quality of these gates using randomized benchmarking \cite{magesan2011scalable}, achieving the effective error rates of $0.00791{\pm}0.00040$, $0.00614{\pm}0.00024$, and $0.00538{\pm}0.00026$ for qubits Q$10$, Q$13$, and Q$15$, respectively.  Details may be found in the Methods section.

The population of the target qubit (Q$12$) in the $Z$-basis is shown in \fig{maj3_mb4_data} for each input state for the $\slsb{3}$, $\slsb{4}$, $\slsb{5}$, and $\slsb{6}$ functions.  We achieve the ASP of $\majfivefid{}{\pm}\majfivestd{}$ and $\majeightfid{}{\pm}\majeightstd{}$ for the $5$ and $8$ entangling gate $\slsb{3}$ circuits, outperforming the classical ASP of $0.875$. For the $\slsb{4}$, $\slsb{5}$, and $\slsb{6}$ functions we again beat the classical ASPs of $0.8125$, $0.71875$, and $0.671875$, where we achieve ASPs of $\mbfid{}{\pm}\mbstd{}$, $\hxzhfid{}{\pm}\hxzhstd{}$, and $\slsbsixfid{}{\pm}\slsbsixstd{}$, respectively.  Each experiment was run with $8000$ shots and repeated $75$ times to build statistics for experimental error bars.  As expected, the $5$ entangling gate $\slsb{3}$ circuit outperforms the $8$ entangling gate circuit, due to the shorter circuit length and fewer entangling gates.  For the $\slsb{4}$ circuit, due to the connectivity restrictions of the device, the logical states of qubits Q$7$ and Q$10$ are swapped twice during the execution of the circuit to interact with the computational qubit Q$12$. Similarly, for the $\slsb{5}$ circuit, the states of Q$7$ is swapped with Q$10$ twice and the states of Q$13$ is swapped with Q$14$ twice. The $\slsb{6}$ follows the same swapping scheme as the $\slsb{5}$ circuit with the addition of swapping Q$15$ and Q$18$ twice.

\section{Conclusion}
In this paper, we established the theoretical advantage of a certain kind of space-restricted quantum computations over analogously defined space-restricted classical computations and demonstrated it experimentally.  Our experiment results in statistics that cannot be reproduced classically.  Specifically, we demonstrated the calculation of $3$-, $4$-, $5$-, and $6$-bit symmetric functions, relying on $1$-qubit limited-space computations, with algorithmic success probabilities of $\majfivefid{}{\pm}\majfivestd{}$, $\mbfid{}{\pm}\mbstd{}$, $\hxzhfid{}{\pm}\hxzhstd{}$, and $\slsbsixfid{}{\pm}\slsbsixstd$ beating the best possible classical statistics of $0.875$, $0.8215$, $0.71875$, and $0.671875$ respectively.  The set of functions computable by $1$-qubit limited-space quantum computations but not $1$-bit limited-space classical computations considered includes functions such as the individual components (such as the Second Least Significant Bit, $\slsb{}$) of the Hamming weight/popcount, a popular classical processor instruction, that is furthermore utilized in fault-tolerant implementations of certain Hamiltonian dynamics algorithms \cite{nam2019low, kivlichan2020improved}.  Our study motivates further development of space-restricted computations unique to quantum computers and an in-depth investigation of the space-time tradeoffs that appear to manifest very differently in the quantum compared to the classical worlds.


\afterpage{\clearpage}

\vspace{4mm}
\textbf{Acknowledgements.} We thank Naoki Kanazawa and Edward Chen for experimental contributions and Jay M. Gambetta for discussions. S.B.~and T.Y.~are partially supported by the IBM Research Frontiers Institute.






\section*{\normalsize Methods}

\section*{Theory}

\noindent{\bf Classical vs quantum limited-space computations.} Here we consider how many classical bits are required to be traded for a single quantum bit to equalize the complexities of classical and quantum limited-space computations, focusing on the Majority function.  We show evidence that the number of classical bits needed to replace a single quantum bit may be $3$ or higher.

Firstly, the computation of the n-bit Majority function with a branching program of width $3$ is conjectured to require a superpolynomial length branching program \cite{razborov1991lower}, \cite[page 432]{wegener1987complexity}.  This directly translates to a $2$-bit ($2 \,{=}\, \lceil \log_2(3)\rceil$) superpolynomial size limited-space classical circuit.  Thus, the comparison of perfect 1-qubit quantum computations to $2$-bit classical computations suggests a quantum superpolynomial advantage in the gate count.  

With $\lceil \log_2(5) \rceil\,{=}\,3$ bits, one may apply Barrington’s theorem to obtain a polynomial-size classical limited-space circuit \cite{barrington1989bounded}.  However, the best limited-space circuit we were able to find based on constructions directly available in the literature has cost $O(n^{10.6})$; it is obtained by employing $5.3{\cdot}\log(n)$-depth Majority Boolean circuit of Valiant \cite{valiant1984short} on top of Barrington’s theorem \cite[Theorem 1]{barrington1989bounded}.  A slightly better construction features cost $O(n^{5.42})$ and requires a little work.  First, compute the input weight $W$ of the $n$-bit input pattern.  This can be done by a branching program of length $O(n^{5.42})$ based on the depth $\log_{3/2}(n)\,{+}\,O(\log\log(n))$ circuit \cite[Lemma 4]{bravyi2020efficient}.  The $n$-bit Majority can now be obtained by comparing $W$ to the integer-valued constant $\lceil n/2 \rceil$, i.e., by the integer comparator appended to the weight calculation.  To compare $k$-bit numbers $x$ and $y$ (here, $k{=}\log(n)$), it suffices to find the most significant digit of the integer sum $x \,{+}\, \bar{y} \,{+}\, 1$, where $\bar{y}$ is the bitwise negation of $y$ (thus, $\bar{y} \,{=}\, 2^k{-}1{-}y$).  This can be done by a classical adder circuit of depth $\log(k)\,{=}\,\log\log(n)$.  Given the weight, computing the Majority thus adds a polylogarithmic factor \cite{barrington1989bounded} to the length of the branching program and can be discarded.  The total length of the branching program and thus classical limited-space circuit implementing the $n$-bit Majority with $3$ bits of computational space is thus $O(n^{5.42})$.  The comparison of perfect $1$-qubit quantum to $3$-bit classical computations thus offers a polynomial ($n^{5.42} \,{\mapsto}\, n^2$) advantage by quantum computations, to the best of our knowledge.  It is unclear if $3{:}1$ or higher ratio describes true advantage---one where classical and quantum gate counts would be asymptotically equal at a fixed tradeoff ratio $C{:}1$ of bits to qubits.

\noindent{\bf The determinant constraint.} Here we explain why the extra phase factor $i$ in the definition of the quantum limited-space model is unavoidable.  We examine the modified version of the model without the extra phase factor and argue that it is capable of computing only linear functions due to a certain determinant constraint. 

Let  $x\,{\in}\,\{0,1\}^n$ be the input bit string and $V(x)=V_L(x)\cdots V_2(x) V_1(x)$, where each gate $V_i(x)$ is a single-qubit unitary operator that  depends on at most one bit of $x$.  Suppose $V(x)$ evaluates a Boolean function $f(x)$ such that 
\begin{equation}
\label{phase1}
V(x)|b\rangle =|b\oplus f(x)\rangle
\end{equation}
for all $x\,{\in}\, \{0,1\}^n$ and $b\,{\in}\, \{0,1\}$. We claim that this is possible only when $f(x)$ is a linear function.  Indeed, Eq.~(\ref{phase1}) implies that $V(x)\,{=}\,X^{f(x)}$, where $X$ is the Pauli-$X$ operator.  Since $\det(X)\,{=}\,{-}1$, one gets
\[
(-1)^{f(x)} = \det(V(x)) = \prod_{j=1}^L \det(V_j(x)).
\]

Suppose $V_j(x)$ depends on the bit $x_{a(j)}$, where $a(j)\in \{1,2,\ldots,n\}$.  Then $\det(V_j(x))=\exp{[i\alpha_j + i\beta_j x_{a(j)}]}$ for some real-valued coefficients $\alpha_j,\beta_j$. We conclude that 

\begin{align}
\label{phase2}
(-1)^{f(x)} & =\prod_{j=1}^L e^{i\alpha_j + i\beta_j x_{a(j)}} \nonumber \\
&=(-1)^{f(0^n)} \prod_{j=1}^L e^{i\beta_j x_{a(j)}} \nonumber \\
& =(-1)^{f(0^n)} \prod_{p=1}^n \exp{[i\gamma_p x_p]},
\end{align}
where  $\gamma_p$ is the sum of all coefficients $\beta_j$ with $a(j){=}p$.  Let $e^p\,{\in}\,\{0,1\}^n$ be a bit string with a single non-zero at the $p$-th bit.  From Eq.~(\ref{phase2}) with $x{=}e^p$ one gets $e^{i\gamma_p} =(-1)^{f(e^p)+f(0^n)}$ and thus 
\[
f(x) =f(0^n) + \sum_{p=1}^n (f(e^p)+f(0^n))x_p {\pmod 2}
\]
for all $x$. This means that $f(x)$ is a linear function. 

To enable computation of non-linear functions we introduce an extra phase factor in Eq.~(\ref{phase1}) such that
\begin{equation}
\label{phase3}
V(x)|b\rangle =\left\{ \begin{array}{rcl}
|b\rangle &\mbox{if}& f(x)=0,\\
i|b\oplus f(x)\rangle &\mbox{if}& f(x)=1.\\
\end{array}\right. \nonumber
\end{equation}
In other words, $V(x)\,{=}\,(iX)^{f(x)}$.  The determinant constraint no longer applies since $\det(V(x))\,{=}\,1$ for all $x$.

\vspace{3mm}
\noindent{\bf Quantum advantage.}
Here we show that any function computable by the classical 1-bit limited-space model has certain linear features.  This prevents the model from approximating most of the symmetric functions, including $\slsb{n}$ and $\maj{n}$.  We show that maximally nonlinear (bent) functions are among the hardest to approximate for the classical model.  We give examples of bent functions that can be computed on all inputs by short quantum 1-qubit limited-space circuits.  Finally, we argue that the observed quantum advantage is robust against noise.

We start with the classical limited-space model.  Recall that we consider $n$ input bits $x\,{=}\,(x_1,x_2,\ldots, x_n)$ stored in read-only memory and one ancilla bit that serves as a scratchpad.  The ancilla is initialized in the state $0$.  At each computational step, it is allowed to examine a single input bit $x_j$ and apply an arbitrary $1$-bit gate to the ancilla.  This gate may depend on the value of $x_j$.  Such computation can be expressed by a program composed of the elementary gates
\[
\begin{array}{rcl}
\flip & : & \mbox{flip the ancilla},\\
\reset({c}) & :  & \mbox{reset the ancilla to $c$},\\
\flip({j},{b}) & :  &  \mbox{if $x_j\,{=}\,b$ then flip the ancilla},\\
\reset({j},{b},{c})  & : & \mbox{if $x_j\,{=}\,b$ then reset the ancilla to $c$}.\\
\end{array}
\]
Here $1\,{\le}\,j\,{\le}\,n$ and $b,c\,{\in}\, \{0,1\}$ are gate parameters.  The program is said to compute a Boolean function $f(x)$ if the final state of the ancilla is $f(x)$ for all $x\,{\in}\, \{0,1\}^n$.  Let $\Omega_n$ be the set of all Boolean functions $f$  with $n$ input bits that can be computed by such programs.  Note that no restrictions are imposed on the program length.

First, we claim that any function $f\,{\in}\,\Omega_n$ can be computed by a simplified program that contains at most one gate $\reset({j},{b},{c})$ for each $j\,{\in}\, \{1,2,\ldots,n\}$.  Indeed, suppose the instruction $\reset({j},{b'},{c'})$ appears before $\reset({j},{b},{c})$. If $b'{=}b$, then removing  $\reset({j},{b'},{c'})$ does not change the function computed by the program.  Specifically, if $x_j{\ne}b$, then none of the two gates is applied.  If $x_j{=}b$, then both gates are applied but all computations that occurred before $\reset({j},{b},{c})$ are irrelevant since the gate resets the ancilla.  In the remaining case, when $b'{\ne} b$, one can replace $\reset({j},{b'},{c'})$ by $\reset(c')$.  Indeed, if $x_j{=}b$, then all gates preceding $\reset({j},{b},{c})$ can be ignored.  Otherwise, if $x_j{\ne} b$, then $x_j{=}b'$ and thus $\reset({j},{b'},{c'})$ is equivalent to $\reset(c')$.  This proves the claim.

Consider the simplified program discussed above.  Let $k$ be the total number of gates $\reset({j},{b},{c})$.  Choose the order of input variables $x_j$ such that the program has the form
$P_k$, 
$\reset({k},{b_k},{c_k})$, $\ldots$, 
$P_2$, 
$\reset({2},{b_2},{c_2})$,
$P_1$,
$\reset({1},{b_1},{c_1})$,
$P_0$. Here $P_i$ are some programs composed of gates $\flip$, $\reset(c)$, $\flip(j,b)$ only and $b_i,c_i$ are some gate parameters. 
It is crucial that any program composed of gates $\flip$, $\reset(c)$, $\flip(j,b)$ computes a linear function of $x$. Thus the full program computes a function  $f(x)$ that becomes linear if we restrict the inputs $x$ to one of the subsets
\[
M_j = \{ x: \, x_j\,{=}\,b_j \mbox{ and } x_i\,{\ne}\, b_i \mbox{ for $1\,{\le}\,i\,{<}\,j$}\}
\]
with $1\,{\le}\, j\,{\le}\,k$  or
\[
M_{k+1} = \{ x: \, x_i\ne b_i \mbox{ for $1\,{\le}\, i\,{\le}\,k$}\}.
\]
Indeed, if $x\,{\in}\, M_j$, then the tailing gates $\reset({i},{b_i},{c_i})$ with $1\,{\le}\,i\,{<}\,j$ are not applied whereas the gate $\reset({j},{b_j},{c_j})$ is applied.  Thus all computations that occurred before $\reset({j},{b_j},{c_j})$ are irrelevant.  All computations that happen after $\reset({j},{b_j},{c_j})$ are equivalent to the composition $P_{j-1} \cdots P_1 P_0$, which computes a linear function.

Note that restricting the function $f(x)$ to the subset $M_j$ is equivalent to fixing the value for some $j$-tuple of variables.  In particular,  $f\,{\in}\,\Omega_n$ only if $f$ can be made linear by choosing the value of a single variable $x_j$ (restrict $f(x)$ to inputs $x\,{\in}\,M_1$).  One can easily check that making any single variable of the Majority function $\maj{n}$ be constant yields a non-linear function for $n\,{\ge}\,3$. Thus $\maj{n}\,{\notin}\, \Omega_n$.  A similar argument shows that the Second Least Significant Bit function $\slsb{n}\,{\notin}\,\Omega_n$.  Indeed, it follows from the definition that fixing variables in $\slsb{n}$ results in a function with the carry vector equal to the prefix ($0$) or suffix ($1$) of the carry vector of $\slsb{n}$, and neither is a linear function when $n{\geq}3$.

How well can we approximate a given Boolean function $g\,{:}\,\{0,1\}^n\,{\to}\, \{0,1\}$ by a function $f\,{\in}\,\Omega_n$?  Define an approximation ratio $R(g)$ as the maximum fraction of inputs $x\,{\in}\, \{0,1\}^n$ such that $f(x){=}g(x)$, where the maximum is taken over $f\,{\in}\, \Omega_n$.  Note that $R(g)\,{\ge}\, 1/2$ since $\Omega_n$ includes both constant-valued functions.  We claim that 
\begin{equation}
\label{R_upper_bound}
R(g)\le   \frac12\left[ 1 + \hat{g}_{max}\log_2{(4/\hat{g}_{max})}\right],
\end{equation}
where $\hat{g}:\{0,1\}^n\to \mathbb{R}$ is the binary Fourier transform of $g(x)$ defined as $\hat{g}(y)=2^{-n}\sum_{x\in \{0,1\}^n} (-1)^{x\cdot y+ g(x)}$ and 
\[
\hat{g}_{max}=\max_{y\in \{0,1\}^n} |\hat{g}(y)|.
\]
Indeed, let $f\,{\in}\, \Omega_n$ be an optimal approximation to $g$.  Let $l_j\,{:}\,\{0,1\}^n\,{\to}\, \{0,1\}$ be a linear function such that  $f(x)\,{=}\,l_j(x)$ for $x\,{\in}\, M_j$.  Since the set of all $n$-bit strings is the disjoint union of the subsets $M_1,M_2,\ldots,M_{k+1}$, one gets

\begin{align}
\label{eq1}
R(g)
=&  \sum_{j=1}^{k+1} \mathrm{Pr}[x{\in} M_j] \cdot \mathrm{Pr}[l_j(x){=} g(x)|x\in M_j].
\end{align}
Here and below the probability is taken over a random uniform $x\,{\in}\, \{0,1\}^n$. Note that $\mathrm{Pr}[x{\in}M_j]\,{=}\,\frac{1}{2^j}$.  By definition of the binary Fourier transform one has
\[
\mathrm{Pr}[l(x){=}g(x)]\le  \frac12\left( 1+\hat{g}_{max} \right)
\]
for any $n$-bit linear function $l(x)$ and thus

\begin{align}
\label{eq2}
\mathrm{Pr}[l_j(x){=}g(x)|x\in M_j]
\le \frac12\left( 1+2^j \hat{g}_{max}\right).
\end{align}
Here we noted that each Fourier component of the restricted function $g(x)|_{x\in M_j}$ is a linear combination of $2^j$ Fourier components of $\hat{g}(y)$ with coefficients $\pm 1$.  Define $k_0=\min\{k{+}1,\,\log_2(1/\hat{g}_{max})\}$ and split the sum in Eq.~(\ref{eq1}) into two terms: those with $j\,{\le}\, k_0$ and those with $j\,{>}\,k_0$.  Using trivial bound $\mathrm{Pr}[l_j(x){=}g(x)|x{\in}M_j]\,{\le}\, 1$ for $j\,{>}\,k_0$ and the bound in Eq.~(\ref{eq2}) for $j\,{\le}\, k_0$ we arrive at

\begin{align}
\label{eq3}
R(g)\le \frac12\left( 1 + k_0 \hat{g}_{max}\right) + \frac{1}{2^{k_0}}.
\end{align}
Furthermore, the term $\frac{1}{2^{k_0}}$ appears only if $k_0\,{<}\,k{+}1$, in which case $k_0\,{=}\,\log_2(1/\hat{g}_{max})$ and $\frac{1}{2^{k_0}}\,{=}\,\hat{g}_{max}$.  Now Eq.~(\ref{R_upper_bound}) follows from Eq.~(\ref{eq3}).  Let us point out that the bound Eq.~(\ref{R_upper_bound}) is tight up to the logarithmic factor. Indeed, it follows directly from the definitions that $R(g)\ge \frac{1}{2}(1+\hat{g}_{max})$.

It is known \cite[Section 5.3]{o2014analysis} that the binary Fourier coefficients of $\maj{n}$ have magnitude at most $\sqrt{2/\pi n}$.  From Eq.~(\ref{R_upper_bound}), it follows that $R(\maj{n})\le 1/2 + O\left(\log(n)/\sqrt{n}\right)$, approaching $1/2$ for large $n$.  Thus $\maj{n}$ is hard to approximate for the classical limited-space model.

A simple calculation shows that any Fourier coefficient of $\slsb{n}$ has magnitude $1/\sqrt{2}^{n}$ for even $n$ and magnitude $\sqrt{2}/\sqrt{2}^{n}$ for odd $n$.  Thus, from Eq.~(\ref{R_upper_bound}) we obtain $R(\slsb{n})\,{\le}\, 1/2 {+} O(n/\sqrt{2}^n)$.  This shows that $\slsb{n}$ approaches the minimal possible threshold $p{=}1/2$ exponentially fast.  Another example of this type of behavior is given by the inner product function $\ip{n}(x):=\sum_{i=1}^{n/2} x_{2i-1} x_{2i} {\pmod 2}$, defined for even $n$.  One can easily check that any Fourier coefficient of $\ip{n}$ has magnitude $1/\sqrt{2}^{n}$ and thus $R(\mathrm{IP}n)\,{\le}\, 1/2 {+} O(n/\sqrt{2}^n)$.  We note that $\slsb{n}$ and $\mathrm{IP}n$ are examples of bent functions (maximally non-linear), featured prominently in cryptography and certain quantum algorithms~\cite{rotteler2010quantum}.

We numerically computed a pruned lookup table of all functions in $\Omega_n$ for small $n$, establishing $R(\maj{3})\,{=}\,R(\slsb{3})\,{=}\,\frac{7}{8}$, $R(\slsb{4})\,{=}\,\frac{13}{16}$, $R(\slsb{5})\,{=}\,\frac{23}{32}$, $R(\maj{5})\,{=}\,\frac{25}{32}$, and $R(\slsb{6})\,{=}\,\frac{43}{64}$.  We also found that $\slsb{n}$ is the hardest to approximate for small $n$ in the sense that $R(\slsb{n})\,{\le}\,R(f)$ for any symmetric function $f$ over $3$ to $6$ Boolean variables.

The above no-go results can be easily  extended to probabilistic  computations.  Let $g\,{:}\, \{0,1\}^n\,{\to}\, \{0,1\}$ be a fixed Boolean function. Define an approximation ratio $R^*(g)$ as the maximum fraction of inputs $(x,r)\,{\in}\, \{0,1\}^{n+m}$ such that $g(x){=}f(x,r)$, where the maximum is taken over all integers $m\,{\ge}\,0$ and over all functions $f\,{\in}\,\Omega_{n+m}$. Here $r\,{\in}\, \{0,1\}^m$ represents randomness consumed by the algorithm. We claim that $R^*(g)\,{\le}\, R(g)$.  Indeed, let $f_r(x){=}f(x,r)$. Then $f_r\,{\in}\, \Omega_n$ for any fixed $r$.  Thus the fraction of inputs $x$ such that $f_r(x){=}g(x)$ is at most $R(g)$ for any fixed $r$. By linearity, the fraction of inputs $(x,r)$ such that $f(x,r){=}g(x)$ is at most $R(g)$.

We establish quantum advantage by showing that the bent functions $\slsb{n}$ and $\mathrm{IP}n$ can be computed by the quantum 1-qubit limited-space circuits discussed next. 

$\slsb{n}$ with inputs $x_1,x_2,...,x_n$ and output $y$ can be expressed as a quantum circuit $c\hgate\xgate(x_n;y)$ $c\hgate\xgate(x_{n-1};y)...c\hgate\xgate(x_1;y)\,\,\, c\hgate(x_1;y)c\hgate(x_2;y)...c\hgate(x_n;y)$, where $c\hgate\xgate$ and $c\hgate$ are controlled versions of $\hgate\xgate$ and $\hgate$ gates, correspondingly, with control on the first and target on the second qubit.  The transformation applied to the qubit $y$ can be described as the matrix product $\hgate^{|x|} (\xgate\hgate)^{|x|}$.  This product cycles through $8$ matrices, $\idgate, \zgate, -i\ygate, -\xgate, -\idgate, -\zgate, i\ygate,$ and $\xgate$, as $|x|$ grows, resulting in the computational basis measurement pattern given by the prefix of the infinite repeating string $00110011...$.  This describes the behavior of the function $\slsb{n}$---indeed, the $n$th bit of this string counting from zero computes the Second Least Significant Bit of the counter, $n$.  We conclude by observing that the middle two gates can be merged into one, obtaining optimized implementation of the $\slsb{n}$ function with $2n{-}1$ gates,

\vspace{-6mm}
\begin{eqnarray*} 
\slsb{n} = c\hgate\xgate(x_n;y)c\hgate\xgate(x_{n-1};y)...c\hgate\xgate(x_2;y) \\
c\zgate(x1;y) \,\,  c\hgate(x_2;y)c\hgate(x_2;y)...c\hgate(x_n;y). \nonumber
\end{eqnarray*}
$\slsb{n}$ admits true implementation with $4n$ ($4n{-}2$, simplified) gates, based on the formula $(i\xgate)^{\slsb{n}(x)}=\hgate^{|x|}(\hgate\xgate)^{|x|} \pgate^{|x|}(\pgate^\dagger\xgate)^{|x|}$.

$\ip{n}$ with inputs $x_1,x_2,...,x_n$ ($n$ is even) and output $y$ can be obtained by a length $3n/2$ circuit, as follows, $\ip{n}=\mgate(x_1,x_2;y)\mgate(x_3,x_4;y)...\mgate(x_{n-1},x_n;y),$ where $\mgate$ is the Margolus gate \cite[page 183]{nielsen2011quantum} that computes Boolean product of the first two qubits into the third up to a relative phase at the cost of three entangling gates.

Finally, let us discuss noise. Suppose $U$ is a limited-space quantum circuit with $L$ gates computing some Boolean function $f(x)$ on all inputs $x$.  We consider a toy noise model such that the noisy version of $U$ computes $f(x)$ on a fraction  $\frac{1}{2}{\left[1 + (1{-}\varepsilon)^L\right]}$ of inputs $x$ for some constant error-per-gate rate $\varepsilon$. Here we assume that each gate of $U$ fails  independently with probability $\varepsilon$ and thus none of the gates fails with probability $(1-\varepsilon)^L$.  Suppose $f$ is a bent function such as $\slsb{n}$ or $\ip{n}$.  The above shows that  $R(f){\le} 1/2 \,{+}\, O(n2^{-n/2})$.  Thus the noisy quantum circuit achieves a higher approximation ratio than any classical limited-space circuit if $(1{-}\varepsilon)^L\,{\gg}\, n 2^{-n/2}$.  Substituting $L{=}3n/2$ one concludes that the quantum advantage can be established with the help of function $\ip{n}$ for sufficiently large $n$ so long as the error rate stays below a constant threshold, $\varepsilon < 1 {-} \frac{1}{\sqrt[3]{2}} \approx 0.206299\ldots$.  This means that with a large number of qubits, the minimal gate fidelity required to demonstrate a quantum advantage can be very low. 


\vspace{3mm}
\noindent{\bf Quantum Signal Processing.} Here we describe 
a method to compute an $O(n^2)$-gate quantum 1-qubit limited-space circuit for any symmetric Boolean function in classical polynomial time.

Signal processing begins with a simple question \cite{low2016methodology}. \emph{Question 1}: Suppose we fix a positive integer $L$. Can a given unitary $U(\phi)$ be written as
\begin{equation}\label{eq:U}
U(\phi)=R_z(\xi_0)\prod_{j=1}^LR_z(\xi_j)R_x(\phi)R^\dag_z(\xi_j),
\end{equation}
for a selection of real numbers $\xi_j$, $j=0,1,\dots,L$? Notice that this is a question of functional equivalence---one has to construct $U(\phi)$ for \emph{all} values of the ``signal" $\phi$, a real number.

We may always write $U(\phi)\,{=}\,A(\phi)I+iB(\phi)X+iC(\phi)Y+iD(\phi)Z$ for real-valued functions $A$, $B$, $C$, and $D$.  In fact, because $R_x(\phi)=\frac12e^{-i\phi/2}\left(I{+}X\right)+\frac12e^{i\phi/2}\left(I{-}X\right)$, we see that $A$, $B$, $C$, and $D$ are functions of $t=e^{i\phi/2}$.  Indeed, they are Laurent polynomials in $t$ with degree $L$.  For example, $A(t)=\sum_{j=-L}^La_jt^j$.

With this setup, we claim that Question 1 has an affirmative answer if and only if
\begin{enumerate}[(i)]
\item $A(t)^2+B(t)^2+C(t)^2+D(t)^2=1$.
\item $A$, $B$, $C$, and $D$ are Laurent polynomials of degree at most $L$, and at least one has degree $L$.
\item Each $A$, $B$, $C$, and $D$ is an even function if $L$ is even and an odd function if $L$ is odd. 
\item $A(t)$ and $D(t)$ are reciprocal functions, i.e.~$A(t)=A(1/t)$. Similarly, $B(t)$ and $C(t)$ are anti-reciprocal, i.e.~$B(t)=-B(1/t)$.
\end{enumerate}
Reference \cite{haah2019product} contains the proof and an efficient algorithm to find the necessary angles $\xi_j$. Notice that the symmetries imply that the values of $A$, $B$, $C$, and $D$ outside the region $\phi\in[0,\pi)$ are completely dependent on their values inside the region. For instance, $A(\phi)=A(-\phi)=-A(2\pi-\phi)=-A(-2\pi-\phi)$.

Often, signal processing is used to create a desired, complex behavior $U(\phi)$ when one can easily create the simple behavior $R_x(\phi)$. However, in certain situations, one would prefer not to have to specify all four functions $A$, $B$, $C$, and $D$. For our purposes, for example, only the behaviors of $A$ and $B$ matter. Thus, a second, complementary question is the following. \emph{Question 2}: Given $A(t)$ and $B(t)$, do there exist functions $C(t)$ and $D(t)$ such that $A$, $B$, $C$, and $D$ together satisfy the conditions (i-iv)?

Question 2 has an affirmative answer if and only if $A$ and $B$ are Laurent polynomials with the symmetries required by (ii--iv) and $0\,{\le}\, A(\phi)^2{+}B(\phi)^2 \,{\le}\,1$ for all real $\phi$. Moreover, the computation of angles $\xi_j$ remains efficient \cite{haah2019product}.

We take these general principles of QSP and apply them to the computation of a symmetric Boolean function $f:\{0,1\}^n\rightarrow\{0,1\}$. Without loss of generality, we assume $f(0^n)=0$. As described in the main text, we define $\phi_x\,{=}\,\Delta|x|{-}\delta$ for the input bit string $x\,{\in}\,\{0,1\}^n$. In general, if we know nothing more about $f$, then we choose $\Delta=\pi/(n+1)$ and $\delta=0$. This places the points of interest, $\phi_x$ for $|x|=0,1,\dots,n$, in $[0,\pi)$. For specific functions with certain symmetries we can obtain shorter circuits by choosing $\Delta$ and $\delta$ more carefully, as we describe later. In the end, our goal is the construction of $U$ so that $U(\phi_x)$ equals $I$ when $f(x)\,{=}\,0$ and equals $iX$ when $f(x)\,{=}\,1$.

Achieving this goal requires the construction of $A$ and $B$ satisfying conditions (ii--iv), $0\,{\le}\, A(\phi)^2{+}B(\phi)^2\,{\le}\, 1$, and

\vspace{-4mm}
\begin{align}\label{eq:Apoints}
A(\phi_x)&=\bigg\{\begin{array}{ll}
0,&f(x)=1\\
1,&f(x)=0
\end{array},\\\label{eq:Bpoints}
B(\phi_x)&=\bigg\{\begin{array}{ll}
0,&f(x)=0\\
1,&f(x)=1
\end{array}.
\end{align}

\noindent To argue later that $A(\phi)^2{+}B(\phi)^2\,{\le}\,1$ we also require
\begin{equation}\label{eq:derivs}
\frac{d}{d\phi}A(\phi_x)=0\quad\text{and}\quad \frac{d}{d\phi}B(\phi_x)=0,
\end{equation}
for all $x\in\{0,1\}^n$.

Assuming $L$ is odd makes $A$ an odd, reciprocal Laurent polynomial: $a_j{=}a_{-j}$ for odd $j$, and $a_j{=}0$ for even $j$.  Our approach is to use the linear system of $2(n+1)$ equations implied by Eqs.~\eqref{eq:Apoints} and \eqref{eq:derivs} to solve for $a_1,a_3,\dots,a_L$, a total of $(L{+}1)/2$ variables. Because $\frac{d}{d\phi}A(0)=0$, a single equation is automatically satisfied, leaving just $2n{+}1$ to be considered.  Equating the numbers of equations and variables implies we need at most a length $L\,{=}\,4n{+}1$ sequence. A similar argument applies to $B$ and reaches the same conclusion. For the rest of this section, we assume that $A$ and $B$ are the minimum degree Laurent polynomials solving the linear systems.

\begin{figure}[t]
\includegraphics[width=\columnwidth]{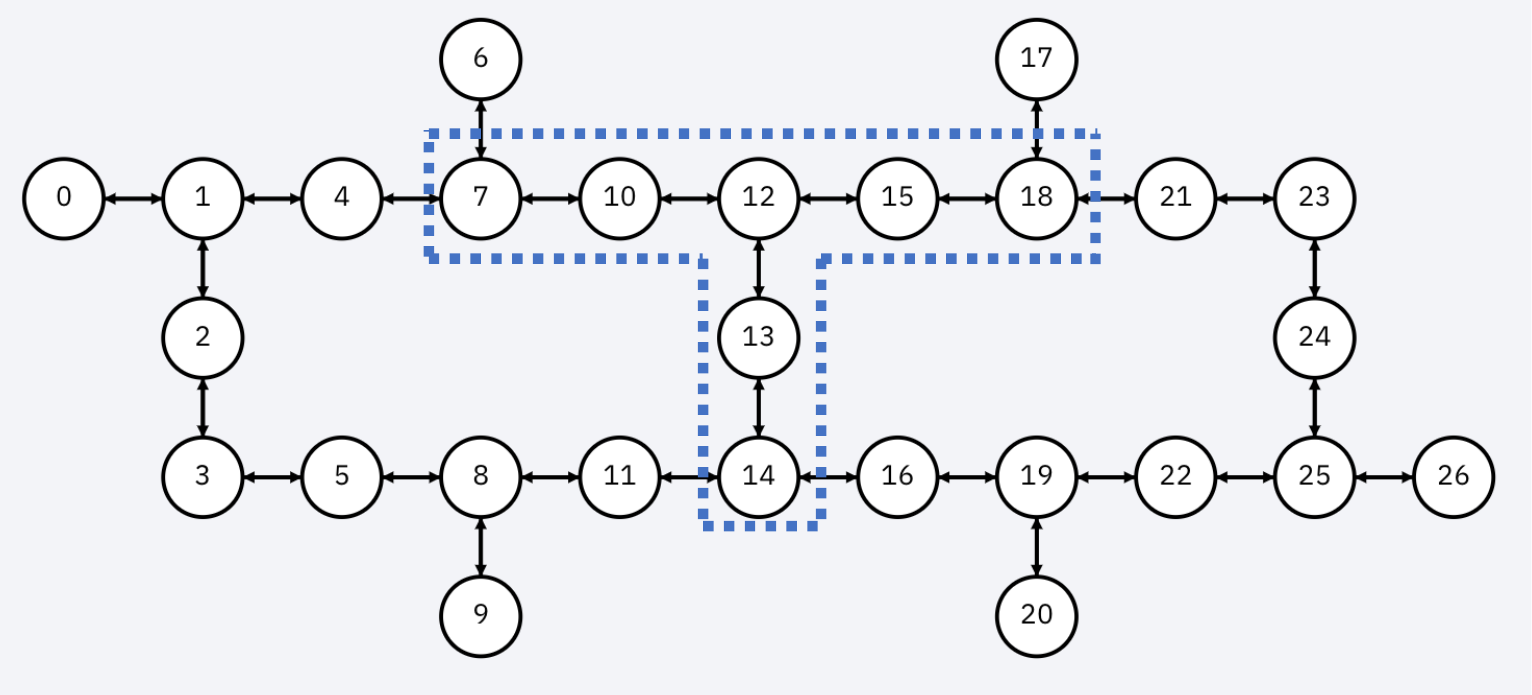}
\caption{Connectivity diagram of \textit{ibmq\_berlin} with the qubits used in the experiment highlighted.}
\label{fig:berlin}
\centering
\end{figure}

Now we show that $P(\phi)\,{=}\,1{-}A(\phi)^2{-}B(\phi)^2$ is never negative. Consider the sum $E(\phi)\,{=}\,A(\phi){+}B(\phi)$, which equals one for all $\phi_x$. Moreover, $E(\phi)$ does not equal one elsewhere because $A$ and $B$ are minimal degree Laurent polynomials satisfying the constraining equations above. Since $\frac{d}{d\phi}E(\phi_x){=}0$ for all $x$ and $E(\phi){<}1$ for some value of $\phi\,{\in}\,[0,\pi]$, we see that $0\,{\le}\, E(\phi)\,{\le}\,1$ for all $\phi\,{\in}\,[0,\pi]$.  By the symmetries of $A(\phi)$ and $B(\phi)$ this implies all the following:

\vspace{-4mm}
\begin{align}
A(\phi){+}B(\phi)&\le1,\text{\space\space} A(\phi)B(\phi)\ge0,\text{\space\space}\text{for }\phi\in[0,\pi],\\\nonumber
A(\phi){-}B(\phi)&\le1,\text{\space\space} A(\phi)B(\phi)\le0,\text{\space\space}\text{for }\phi\in[-\pi,0],\\\nonumber
-A(\phi){+}B(\phi)&\le1,\text{\space\space} A(\phi)B(\phi)\le0,\text{\space\space}\text{for }\phi\in[\pi,2\pi],\\\nonumber
-A(\phi){-}B(\phi)&\le1,\text{\space\space} A(\phi)B(\phi)\ge0,\text{\space\space}\text{for }\phi\in[-2\pi,-\pi].
\end{align}
These together imply that $P(\phi)\,{\ge}\,0$ for all $\phi$. For instance, if $\phi\in[0,\pi]$, then $P(\phi)=1-\left(A(\phi){+}B(\phi)\right)^2+2A(\phi)B(\phi)\ge0$. Similar arguments hold for the other three intervals.

With this, we completed the construction of $U$ to compute the symmetric Boolean function $f$.  At most $4n{+}1$ uses of $R_x(\phi_x)$ are required.  Each of these takes $n$ two-qubit gates to implement, leading to a total of $O(n^2)$ two-qubit gates.  Each application of $R_x(\phi_x)$ is also accompanied by a constant number of single-qubit gates, leading to a total of $O(n)$ single-qubit gates.

\begin{figure}[t]
\includegraphics[width=\columnwidth]{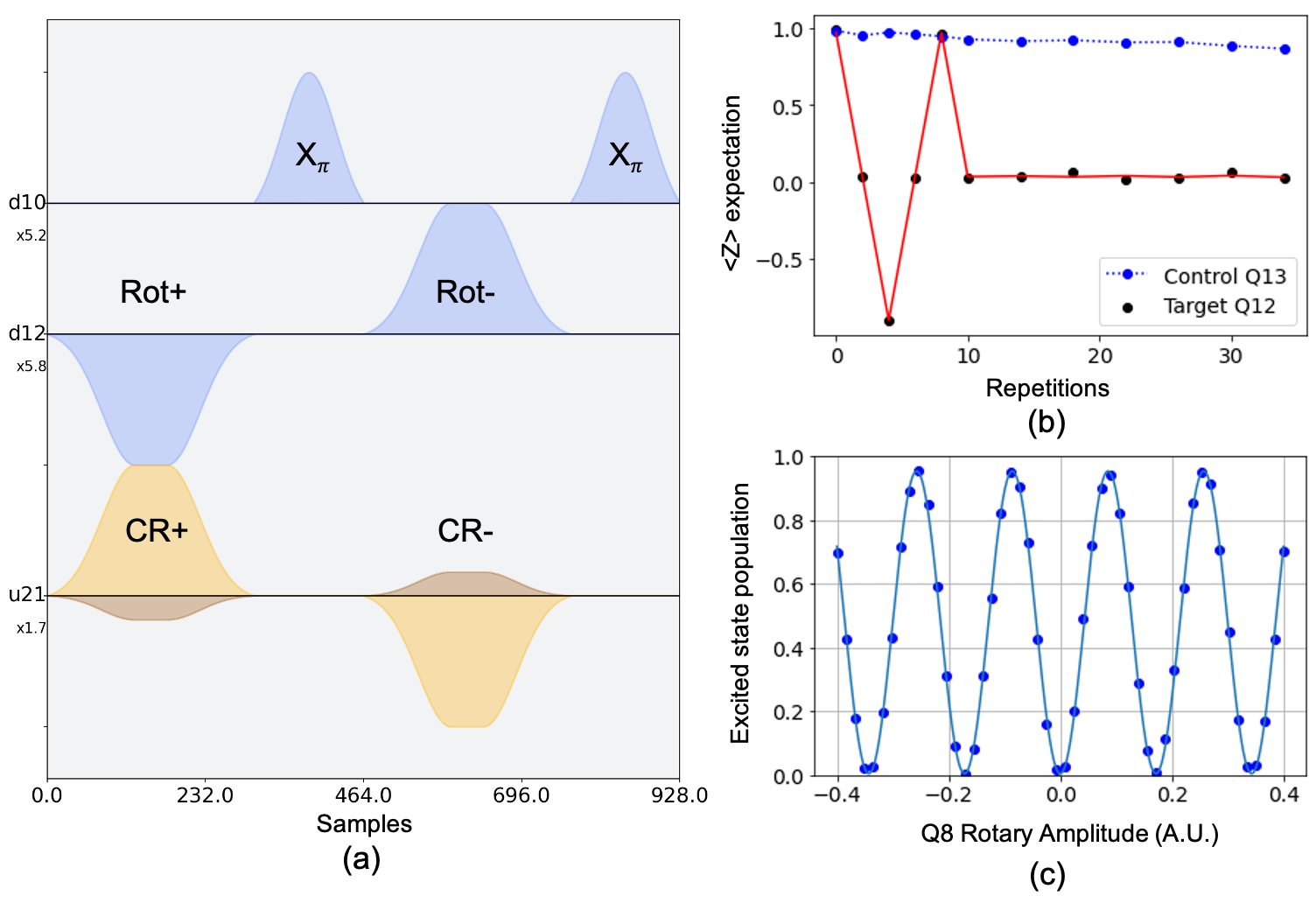}
\caption{(a) Pulse diagram of the echoed CR sequence including the rotary echoes applied to the target qubit. The sampling time is 0.2222 ns per sample. ``d10" and ``d12" denote the drive channels for qubits Q$10$ and Q$12$ respectively, while ``u21" denotes the cross-resonance channel for the control qubit, Q$10$. (b) Fine amplitude calibration of the echoed $ZX_{\pi/4}$ CR pulse sequence for qubits Q$15$ and Q$12$. Initially the $ZX_{\pi/4}$ pulses are applied in repetitions of 2 to ensure a full rotation about the Bloch sphere. At 16 repetitions, the pulses are applied in repetitions of 4 to apply $\pi$ pulses about the Bloch sphere equator in order to amplify amplitude errors. (c) Rabi oscillations of the target qubit used to calibrate a 2$\pi$ rotary echo for Q$12$.}
\label{fig:calibrations}
\centering
\end{figure}

In specializing to $f\,{=}\,\maj{n}$, we obtain shorter sequences by placing the points $\phi_x$ in the range $[-\delta,\pi+\delta]$ with the choices $\Delta\,{=}\,\frac{2\pi}{n+1}$ and $\delta\,{=}\,\frac{\pi}{2}\frac{n-1}{n+1}$. Several equations from Eqs.~\eqref{eq:Apoints} and \eqref{eq:derivs}, those involving points $\phi_x$ outside $[0,\pi)$, become redundant, leaving just $n{+}1$ equations. Thus, we can choose $L\,{=}\,2n{+}1$. We can also set $B(t)\,{=}\,A(-it)$, which means that $B(\phi)$ is $A(\phi)$ shifted by $\pi/2$.

\begin{figure*}[t]
\includegraphics[width=\textwidth]{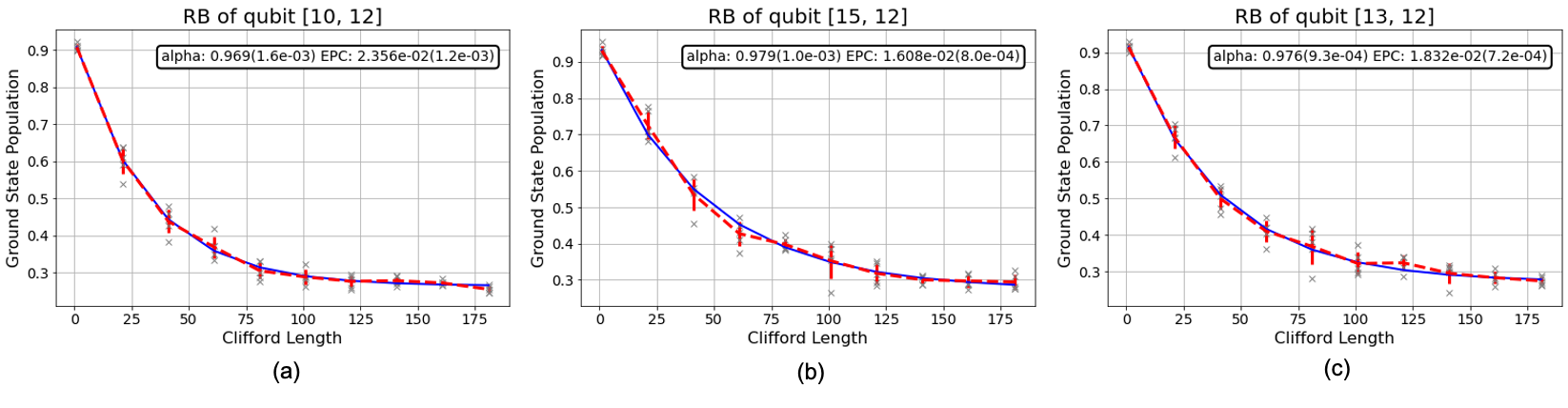}
\caption{Randomized benchmarking for decay curves characterizing the controlled-$R_{x}(\pi/2)$ calibrated in Qiskit Pulse. Pair (Q$10$,\,Q$12$) is shown in (a), (Q$15$,\,Q$12$) in (b), and (Q$13$,\,Q$12$) in (c).  Data is shown in red and the fit is shown in blue.  Error per Clifford (EPC) is shown in each plot and the error per controlled-$R_{x}(\pi/2)$ is taken to be $\frac{1}{3}$EPC.}
\label{fig:rb}
\centering
\end{figure*}

We note that sometimes in the case of $\maj{n}$ not all the derivative constraints, Eq.~\eqref{eq:derivs}, are necessary to guarantee $P(\phi)\,{\ge}\,0$. For $n{=}3$ and $n{=}5$ we are able to find suitable $A$ and $B$ by only requiring zero derivatives where the functions evaluate to one---a feat impossible for $n{=}7$.  Moreover, significant simplification may occur when it happens that the consecutive angles $\xi_j$ and $\xi_{j+1}$ are equal.  In this case, subsequent $R_z$ gates in Eq.~\eqref{eq:U} cancel and two $R_x(\phi_x)$ rotations can be combined into a $R_x(2\phi_x)$ gate, which needs just $n$ two-qubit gates, rather than $2n$, to implement.  Hence, we obtain smaller than anticipated circuits---a circuit with $9$ two-qubit gates for $\maj{3}$ and a circuit with $25$ two-qubit gates for $\maj{5}$.

\section*{Experiments}

Our experiments are executed on qubits Q$7$, Q$10$, Q$12$, Q$13$, Q$14$, Q$15$, and Q$18$ of \textit{ibmq\_berlin}, a 27-qubit heavy-hexagonal lattice device (\fig{berlin}). Qubit Q$12$ is used as the scrap space while the rest serve as the input qubits.  The measured qubit parameters are shown in \tab{qubit_params} below.

\begin{table}[h]
\begin{tabular}{ |p{1.2cm}||p{1.5cm}|p{1.5cm}|p{1.5cm}|p{2cm}|}
 \hline
 Qubit &T1 ($\mu$s) &T2 ($\mu$s) &Readout Error (\%) & $\cnotgate$ (Q$_i$,Q$_j$) Error (\%)\\
 \hline
 Q7 & 78.93 & 67.09 & 1.10 & (7,10): 1.287 \\
 Q10 & 122.27 & 133.38 & 3.38 & (10,12): 0.867\\
 Q12 & 69.05 & 107.34 & 1.86 & - \\
 Q13 & 88.30 & 91.43 & 1.80 & (13,12): 0.610 \\
 Q14 & 110.55 & 36.67 & 1.90 & (14,13): 1.939\\
 Q15 & 128.77 & 137.16 & 2.01 & (15,12): 1.361\\
 Q18 & 71.44 & 65.51 & 2.47 & (18,15): 2.674\\
 \hline
 \end{tabular}
 \caption{Qubit parameters and $\cnotgate$ gate errors for the qubits used in the experiment. For the $\cnotgate$ gate errors, values are measured daily utilizing randomized benchmarking.  Parameters are representative of the system over time.}
 \label{tab:qubit_params}
 \end{table}

Single-qubit gates within our circuits are implemented as standard Qiskit gates comprised of the sequences of $X_{\pi/2}$ rotations and virtual $Z$ rotations \cite{mckay2017}.  With these two operations, any single-qubit rotation can be realized with a maximum of two $X_{\pi/2}$ gates. $X_{\pi/2}$ are implemented applying a $35.5$ ns Gaussian microwave pulse at the qubit's frequency with a Gaussian derivative shaped pulse added to the quadrature phase to reduce leakage out of the computational space. $Z$ rotations are implemented virtually as frame changes within the software by simply shifting the microwave phase of each subsequent microwave pulse by the desired angle of rotation. As a result, single-qubit $Z$ rotations are implemented with essentially no error and they take no time.

The standard two-qubit $\cnotgate$ gates are implemented using an echoed cross-resonance (CR) drive in which the control qubit is irradiated with a Gaussian-square microwave tone at the target qubit's frequency (\fig{calibrations} (a)).  The standard echoed CR pulse sequence is as follows: a positive CR tone is applied followed by a $\pi$ pulse. Then a negative CR tone is applied followed by a second $\pi$ pulse. The resulting net rotation is $ZX_{\pi/2}$ for the standard $\cnotgate$ gate.  A resonant 2$\pi$ rotary echo is applied synchronously to the target qubit to minimize cross-talk and unwanted Hamiltonian terms.  The phase of the CR drive is calibrated such that the primary interaction term is $ZX$, relative to the target qubit's frame.  Standard single-qubit and two-qubit gates are auto-calibrated each day. The overall rotation applied is a function of the CR tone amplitude and duration.

To implement custom controlled-$R_{x}(\pi/2)$ gate and the locally equivalent controlled-$\hgate \xgate$ gate we modify the existing $\cnotgate$ parameters using Qiskit Pulse. The width of the Gaussian-square CR tone is reduced by roughly half and the amplitude is recalibrated such that the amplitude of the custom pulse is roughly equal to the standard CR pulse. In addition, the rotary echo pulse is recalibrated on the target qubit to achieve a 2$\pi$ rotation within the same duration as the CR tone (\fig{calibrations} (c)).  As the standard $\cnotgate$ gate CR amplitudes are already optimized to minimize the gate error, our strategy is to maintain this amplitude for the custom CR pulse to minimize gate errors.  The net rotation of the custom pulse is thus $ZX_{\pi/4}$, which is equivalent to the custom gates we implement up to the single-qubit gate corrections. 

We calibrate the custom CR pulse amplitude by applying repeated echoed $ZX_{\pi/4}$ pulses with the rotary echo and reading out the target qubit along the Bloch sphere equator (\fig{calibrations} (b)). In this way, amplitude miscalibrations are amplified and can be corrected.  Three CR tones are calibrated this way corresponding to the three qubits coupled to the computational qubit: (Q$10$,\,Q$12$), (Q$13$,\,Q$12$), and (Q$15$,\,Q$12$).

Once the custom two-qubit gates are calibrated, we measure the error per gate using randomized benchmarking (\fig{rb}).  We substitute two controlled-$R_{x}(\pi/2)$ gates for each required $\cnotgate$ in the Clifford sequence and measure the decay of the $\ket{00}$ population as a function of Clifford gates.  From the decay rate, we can extract an error per Clifford gate (EPC).  Knowing each Clifford is on average 1.5 $\cnotgate$s, and therefore 3 controlled-$R_{x}(\pi/2)$ gates, we can approximate the error rate of the controlled-$R_{x}(\pi/2)$ as $\frac{1}{3}$EPC in the limit of small errors.  We extract the controlled-$R_{x}(\pi/2)$ error rates of $0.00791{\pm}0.00040$, $0.00614{\pm}0.00024$, and $0.00538{\pm}0.00026$ for qubit pairs (Q$10$,\,Q$12$), (Q$13$,\,Q$12$), and (Q$15$,\,Q$12$) respectively.  Since the dominant error comes from the two-qubit interaction, the error rates of the locally equivalent controlled-$\hgate \xgate$ is comparable to the controlled-$R_{x}(\pi/2)$ gates.

\end{document}